\documentclass[usegraphicx]{mn2e}
\title[Origin of field ZAMS stars in the solar neighborhood]{
        Field ZAMS stars in the solar neighborhood:\\
	Where have they come from ?
        \thanks{Based on observations collected at the 
        European Southern Observatory, Chile 
        (ESO N$^o$ 62.I-0650, 66.D-0159(A), 67.D-0236(A))}
}

\author[R. Wichmann \& J.H.M.M. Schmitt]{
        R. Wichmann,$^1$\thanks{Visiting Astronomer, 
                           Kitt Peak National Observatory,
                           National Optical Astronomy Observatory,
                           which is operated by the Association of 
                           Universities for Research in Astronomy, Inc. (AURA)
                           under cooperative agreement with the National 
                           Science Foundation.}
        \and
         J.H.M.M. Schmitt$^1$\\
        $^1$Hamburger Sternwarte, Gojenbergsweg 112, D-21029 Hamburg, Germany
}

\begin{document}

\maketitle

\begin{abstract}
	In the course of an all-sky survey for young stars in the solar
	neighborhood we have found a tight kinematic group of ten
	F-G type zero-age main sequence stars in the field. 
	Here we discuss the origin of these stars.
	Backtracking the space motions of these stars we argue that 
        likely candidates for the parent association are
	the Perseus OB3 (Per OB3), Upper Centaurus-Lupus (UCL), and
	Lower Centaurus-Crux (LCC) associations, and
	that we are
        witnessing the ongoing diffusion of (at least one of) 
	these associations into the field.
\end{abstract}


%

\section{Introduction}

    We have recently performed an all-sky survey for nearby young
    stars, the results of which are published in Wichmann et al. 
    \cite{Wi02}.
    The candidate sample for this survey was obtained by cross-correlating
    the ROSAT All-Sky Survey (RASS) with the TYCHO catalogue and selecting
    stars with ${\rm B} - {\rm V} > 0.54$ (i.e. F8 and later) with
    (TYCHO) parallax errors better than $3.5\,\sigma$. In addition,
    stars very far above the the main sequence (giants) and very far
    below (erroneous parallaxes) were discarded, resulting in a total
    sample size of 754 stars. Because of these selection procedures,
    the majority of our candidate stars are located within 60\,pc of the Sun.

    The spectroscopic follow-up investigations of 748 stars out of 
    this sample resulted in the detection of 10 G-type stars
    (HD\,13183, HD\,35850, HD\,36869, HD\,43989, HD\,49855, HD\,77407, 
    HD\,105070, HD\,129333, HD\,171488, and HD\,202917),
    which share
    the following properties: 
    (1) They all have lithium equivalent
    widths W$_{\rm Li}$ in excess of the Pleiades upper limit of 
    their respective spectral type; 
    (2) They show very high X-ray activity 
    ($\log{L_{\rm X}/L_{\rm bol}} = -3.49 \pm 0.07$); and 
    (3) They show a
    very narrow distribution of their galactic velocity components 
    ($U = -1.0 \pm 2.5$, $V = -16 \pm 3.4$, $W = 1.7 \pm 3.1$).

    All of these observational indicators strongly suggest that these
    stars are quite young: The observed excess lithium, the high X-ray activity
    and the narrow velocity dispersion 
    (cf., Wielen 1977) suggest an age
    no more than about 50 Myrs for these stars. 
    While the youth of some of these stars has 
    been known prior to our study, the fact that these stars do form a very
    tight kinematic group, despite their being distributed
    more or less uniformly over the whole sky with distances
    of up to about 100\,pc from each other, has remained unnoticed so far.
    For a detailed discussion of our survey and these 10 ZAMS 
    stars in our sample we
    refer to our main survey paper (Wichmann et al.~2003).

    We note specifically 
    that this
    kinematic group of stars was identified on the basis of their
    lithium equivalent widths, i.e. without applying any kinematic
    criteria.  In principle these stars -- now distributed over the whole
    sky -- could have formed in an unrelated fashion with a chance
    alignment of their current velocity vectors. However, we feel that 
    a far more natural explanation for this tight kinematic group of
    10 ZAMS stars found in our study is to assume that they have
    some common origin. 
    This begs the question: where do they come from ?
    
\section{The origin of the ZAMS sample}

    Any reasonable candidate for the formation of these stars should be
    close to the Sun, because the youth of these stars and their small
    velocities of a few km\,s$^{-1}$ w.r.t. the local standard of 
    rest (LSR) shows
    that they cannot have moved very far from their birthplace(s). At the same
    time, they must originate from a 
    large star forming region (SFR),
    since an extrapolation based on a normal initial mass function 
    (IMF; c.f. Kroupa et al~1993) shows that these 10 stars
    correspond to a total mass of about 120\,$M_{\odot}$. 

    Furthermore,
    the tight velocity distribution of our ten stars shows that they
    cannot have been ejected by three-body interactions, rather they must
    have slowly dispersed from their birth regions. Thus, any likely
    candidate should show similar galactic velocities, and its
    trajectory should intersect those of the 10 ZAMS stars at some point
    at most $\approx 50$\,Myr in the past. 

    We can safely rule out the
    Taurus-Auriga SFR because of the grossly inconsistent velocity, as well as
    the Pleiades because they are too old.
    After reviewing all known star forming regions close to the Sun, we
    find that the best candidates for the origin of our ten ZAMS stars
    are the following three associations: Perseus OB3 ($\alpha$\,Persei), 
    Lower Centaurus Crux (LCC), and
    Upper Centaurus Lupus (UCL).

%

\section{Data}

    The current position of the centres of the Per OB3, LCC, and UCL 
    associations are based on the membership lists by 
    de Zeeuw et al. \cite{deZeeuw99}. 
    These membership lists were
    obtained from the HIPPARCOS catalogue 
    by applying a combination of two different methods. The first is 
    a modern implementation of the convergent 
    point method as described by de Bruijne \cite{dB99}, while the second 
    is a kinematic member selection method developed by
    Hoogerwerf \& Aguilar \cite{HA99}. This kinematic method 
    also uses the HIPPARCOS parallax
    information, which is neglected in the convergent point method.
    Only members confirmed by both methods were retained.

    Per OB3 is located at a distance of
    about 180\,pc and comprises 30 OB stars.
    Its age is estimated to $\sim 50$\,Myr (de Zeeuw et al.~1999).
    The LCC/UCL associations (together
    with a third, the Upper Scorpius association) are subgroups of the
    Sco OB2 association. For this entire complex, Blauuw \cite{Bl64b}
    estimated an expansion age of $\approx 20$\,Myr.
    The LCC is at a distance of $\approx 120$\,pc and comprises 36-42 OB stars,
    while the UCL is $\approx  140$\,pc distant and has 58-66 OB-type members
    (Ma\'{\i}z-Apell\'aniz~2001, de Zeeuw et al.~1999, 
    Hoogerwerf et al.~2001).
    From Monte-Carlo simulations, de Zeeuw et al. \cite{deZeeuw99} estimate
    that about 10 out of 79  members of Per OB3, 
    some 50 out of 221 members of UCL, and some 30 out of 180
    members of LCC might be interlopers, i.e. field stars unrelated to
    the association. 

    The most recent determination of the velocities of the 
    Sco-Cen sub-groups,
    using HIPPARCOS data as well as radial velocities from various sources
    (listed in Asiain et al.~1999a), has been carried out by
    Asiain et al. \cite{As99b}. The space velocities as listed in their 
    Table 3 were used for this work.
    The current spatial extent of the LCC and UCL associations can 
    be aproximated
    by a Gaussian with $\sigma \approx 25-30$\,pc 
    (Ben\'{\i}tez et al.~2002).
    The nuclear ages as deduced by de Mamajek et al. \cite{Ma02} are
    15-17\,Myr for the LCC, and 16-18\,Myr for the UCL.

    The turnoff age of Per OB3 is estimated to about 50\,Myr 
    by de Zeeuw et al. \cite{deZeeuw99}, and the spatial extent
    to about $3^0 \times 3^0$ (corresponding to
    a radius of about 4.6\,pc),
    with a halo of 10$^o$ (31\,pc). 
    The radial velocity of Per OB3 (-1\,km\,s$^{-1}$)
    determined by de Zeeuw et al. \cite{deZeeuw99} is the median velocity
    compiled from the HIPPARCOS Input Catalogue. 
    Table~\ref{TabSco}
    summarizes the relevant data on the associations considered here.

   \begin{table}
      \caption[]{
	Data for UCL, LCC, and Per OB3. 
	For the definition of the ($\xi$, $\eta$, $\zeta$)
        coordinate system see text. U, V, and W are the
	galactic velocity components (expressed in the LSR)
	towards the galactic centre, the
	direction of rotation, and the galactic pole, respectively (i.e.
	${\rm U} = -\dot{\xi}$, ${\rm V} = \dot{\eta}$, and
	${\rm W} = \dot{\zeta}$). For LCC/UCL, ages are nuclear ages 
	from de Mamajek et al.
	\cite{Ma02}, positions are from Ma\'{\i}z-Apell\'aniz \cite{MA01}, 
	and velocities from 
	Asiain et al. \cite{As99b}; all data for Per OB3 are from
	de Zeeuw et al. \cite{deZeeuw99}.
        }
        \label{TabSco}
        \begin{tabular}{lcrrrrrr}
  \noalign{\smallskip}\hline
         &   Age         & $\xi$ & $\eta$ & $\zeta$ & U    & V         & W \\
  \noalign{\smallskip}\hline
Subgroup & 
(Myr)  & 
\multicolumn{3}{l}{  (pc)} & 
\multicolumn{3}{l}{  (km\,s$^{-1}$)} \\
  \noalign{\smallskip}\hline
LCC      &  15-17        &  -62  & -100 & 37   & -4.9      & -15.6     &  1.2\\
UCL      &  16-18        & -119  &  -67 & 58   & -8.3      & -15.4     &  2.8\\
Per OB3  &  $\approx 50$ &  149  &  100 &  9   &  4.4      & -19.8     & -0.2\\
  \noalign{\smallskip}\hline
\end{tabular}
\end{table}

%

\section{Motion backtracking}

\subsection{Epicyclic approximation}

    We introduce
    a coordinate system ($\xi$, $\eta$, $\zeta$) centered on 
    the projection of the Sun's position on the Galactic plane 
    (we use $\zeta_{\odot} = 27$\,pc from Chen et al.~2001) 
    and in a circular orbit around the galactic centre with
    angular velocity $\omega_c$ and radius $\varpi_0$. The $\xi$-axis points
    away from the galactic centre, the $\eta$-axis points in the 
    direction of rotation, and the $\zeta$-axis toward the galactic pole.

    In this coordinate system, the equations of motion in the ($\xi$, $\eta$) 
    plane are:
    \begin{equation}
    \label{EOM}
    \begin{array}{rcl}
    \ddot{\xi} - 2 \omega_c \dot{\eta} - 4 \omega_c A \xi & = & 0, \\
    \ddot{\eta} + 2 \omega_c \dot{\xi}                        & = & 0 \\
    \end{array}
    \end{equation}
    if the stars move in an axisymmetric central potential and
    $\xi \ll \varpi_0$, $\eta \ll \varpi_0$ ($A$ is Oorts constant). 
    I.e. Eqn.~\ref{EOM} neglects deviations
    from an axisymmetric potential due to local density variations, and
    is valid for small eccentricities only 
    (which is true for our 10 ZAMS stars). The derivation of Eqn.~\ref{EOM}
    is given in Chapt.~5 of Chandrasekhar \cite{Ch42}, although these
    equations of motion were already stated -- without proof -- by 
    Lindblad \cite{Lb36}.

    Following Chandrasekhar \cite{Ch42}, the general solution of 
    Eqn.~\ref{EOM} can be written as:
    \begin{equation}
    \label{EOS0}
    \begin{array}{rcl}
    \xi & = & \xi_{11} \cos{q_1 t} + \xi_{21} \sin{q_1 t} + C_1 t + C_2,\\
    \eta  & = & \lambda_1(\xi_{11} \sin{q_1 t} - \xi_{21} \cos{q_1 t}) + C_3 t + C_4,\\
    \end{array}
    \end{equation}
    where $t$ is the time, $\xi_{11}$, $\xi_{12}$, 
    $C_{1}$, $C_{2}$, $C_{3}$, $C_{4}$ 
    are integration constants to be fixed by the
    initial conditions, 
    $ q_1 = 2\sqrt{\omega_c(\omega_c - A)}, \quad {\rm and} \quad
       \lambda_1 = - \sqrt{\omega_c /(\omega_c - A)}
    $
    ($q_1$ is called the {\it epicyclic frequency}, and often also
    denoted by $\kappa$).
    From Eqn.~\ref{EOM} one can see that $C_{1} \equiv 0$, 
    $C_{3} = - 2 A C_{2}$, and thus
    Eqn.~\ref{EOS0} can be rewritten as: 
    \begin{equation}
    \label{EOS1}
    \begin{array}{rcl}
    \xi & = & \xi_{11} \cos{q_1 t} + \xi_{21} \sin{q_1 t} + \lambda \xi_{12},\\
    \eta  & = & \lambda_1(\xi_{11} \sin{q_1 t} - \xi_{21} \cos{q_1 t}) + \lambda_2 (\xi_{12}t - \xi_{22}),\\
    \end{array}
    \end{equation}
    with integration constants $\xi_{21}$, $\xi_{22}$, and 
    $\lambda = 1/[2(\omega_c - A)], \quad \lambda_2 = - A/(\omega_c - A), $
    which is the solution stated by Lindblad \cite{Lb37}.

    This solution describes an elliptical motion with the centre of the
    ellipse (of axis ratio 1/$\lambda_1$) 
    located at $\xi =  - \lambda \xi_{12}$ and moving parallel
    to the $\eta$-axis with a velocity of $\lambda_2 \xi_{12}$. For this
    reason, the approximation presented above is known as 
    {\it epicyclic} approximation. If the initial values $\xi_0$, $\eta_0$, 
    $\dot{\xi}_0$, and $\dot{\eta}_0$ at $t = 0$ are known, the integration
    constants can be evaluated to:
    \[
	\xi_{21} = \frac{\dot{\xi}_0}{q_1}, \quad
	\xi_{22} = \frac{1}{\lambda_2}\left( \frac{\lambda_1 \dot{\xi}_0}{q_1} + \eta_0 \right),
    \]
    \[
	\xi_{12} = \dot{\eta}_0 - \lambda_1 q_1 \xi_0 \lambda, \quad 
        \xi_{11} = \lambda_2 \xi_0 - \lambda \dot{\eta}_0.
    \]

   \begin{figure}
   \centering
   \includegraphics[width=6.4cm,angle=270]{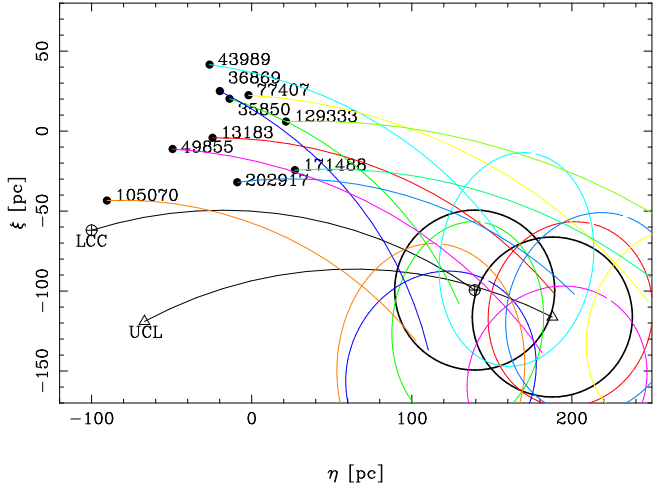}
      \caption{Current location of the 10 field ZAMS stars (filled
	circles, annotated with HD numbers) and their trajectories 
        backwards to $t = -15$\,Myr,
	with 1\,$\sigma$ error ellipses for the
        positions at $t = -15$\,Myr
	(see text). Also plotted are the trajectories of the 
	LCC/UCL
	OB associations, with circles centered on their positions at
	$t = -15$\,Myr indicating their 2\,$\sigma$ radii.
	}
         \label{FigOne}
   \end{figure}

   \begin{figure}
   \centering
   \includegraphics[width=6.4cm,angle=270]{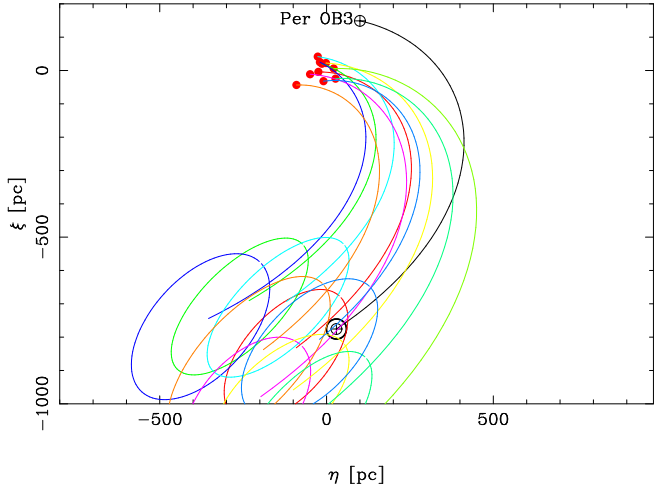}
      \caption{Same as Fig.~\ref{FigOne}, but for Per OB3 at $t = -50$\,Myr.
	}
         \label{FigTwo}
   \end{figure}

\subsection{Error propagation}

    Any error in the velocity will lead to increasing position errors
    with time when
    tracing the trajectory backwards, and must be accounted for carefully.
    In our error budget, we include the errors in the proper
    motions as given in the HIPPARCOS catalogue, as well as errors 
    in the radial
    velocities measured by us ($ \approx 1.5$\,km\,s$^{-1}$).

    The formalism presented above does not account for local 
    density fluctuations. We treat this effect as
    an additional error, leading to random velocity variations and 
    accordingly to a diffusion of the orbits. The rms variations
    of the galactic velocity components as a function of time have
    been studied empirically by Wielen \cite{Wi77}.
    Errors in right ascention and declination are small, and thus we
    neglect them, while errors in the parallax are taken into account.
    For the constants in Eq.~\ref{EOS1}, we use $\omega_c = 27.19$\,kpc/Myr
    and $A = 14.82$\,kpc/Myr (Feast \& Whitelock~1997).

    We then proceed by performing, individually for each star, 
    a Monte-Carlo simulation to determine an error ellipse for the
    position at $t = -15$\,Myr (for LCC) 
    and $t = -50$\,Myr for $\alpha$ Per. We
    draw random proper motions and radial velocities from a 
    Gaussian distribution with mean value and dispersion corresponding to the
    measured values and their errors,
    compute the galactic UV velocities, and add random Gaussian noise according
    to Table 2 of Wielen \cite{Wi77} to account for 
    local density fluctuations. We then compute the backward trajectory
    for 15 Myr. After 10\,000 trials, we finally fit a 2$\sigma$
    error ellipse around all trial positions (at $t = -15$\,Myr and 
    $t = -50$\,Myr respectively). This
    error ellipse is computed by the following procedure: first, the centre
    of the distribution of positions is computed. A straight
    line is fitted to the distribution and adopted as major
    axis. The axis ratio is determined by the ratio of the dispersions
    along the major and minor axes. Finally, the length of the major axis
    is chosen such that 95 percent of the positions are located within
    the ellipse.

    In Fig.~\ref{FigOne},\ref{FigTwo} we show the resulting error ellipses, 
    as well as the trajectories for the measured velocities.
    In Fig.~\ref{FigOne}, we also plot the trajectories
    for the LCC and UCL associations, and 
    the 2$\sigma$ radii of 50\,pc
    (Ben\'{\i}tez et al.~2002) for the LCC/UCL
    associations, centered at the position at $t = -15$\,Myr.
    In Fig.~\ref{FigTwo}, the trajectory for Per OB3 is shown, along with its
    position at $t = -50$\,Myr.
    The radius for Per OB3 corresponds to the 10$^0$ (30\,pc) 
    halo.

\subsection{Motion perpendicular to the plane}

   In the approximation of a flattened, homogeneous mass distribution, 
   the equation of motion in the $\zeta$-direction is
\begin{equation}
   \ddot{\zeta} + \frac{\partial^2 \Psi}{\partial \zeta^2} \zeta = 0, 
\end{equation}
   where $\Psi$ is the gravitational potential 
   (c.f. Chandrasekhar~1942). 
   This is solved by
\begin{equation}
   \zeta = \zeta_1 \cos q_3 t + \zeta_2 \sin q_3 t - C_5 t + C_6,
\end{equation}
   with $q_3 = \sqrt{4 \pi \rm{G} \rho_0}$ and integration constants
   $C_5 = C_6 = 0$, $\zeta_1 = \dot{\zeta}_0$, and $\zeta_2 = \zeta_0$.
   Here, $G$ is the gravitational constant, $\rho_0$ the volume density
   in the galactic plane, and $\zeta_0, \dot{\zeta}_0$ are the initial
   conditions at $t = 0$. We neglect the (small) 
   term $2(2A - \omega_c)\omega_c$
   in the more general expression for $q_3$ (c.f. Mihalas \& Routy~1968).

%

\section{Discussion}

    We are aware that some of these 10 stars have been assigned to several
    small star forming regions. In particular, 
    HD\,13183 is claimed by Torres et al.
    \cite{Torres00} as a probable member of the Horologium association, 
    and HD\,35850
    by Zuckerman et al. \cite{Zuckerman01} as a member of the 
    $\beta$ Pic association.
 
    However, considering the very small observed
    velocity dispersion, we consider it unlikely that these 
    stars have formed in different and unrelated star forming regions.
    We do not think that this is in contradiction to the results of
    Zuckerman et al. \cite{Zuckerman01} or Torres et al.
    \cite{Torres00}, because membership assignment in an association 
    is invariably based
    on statistical criteria which always have some error of the second kind
    (i.e. accepting membership for a non-member).

    Also, we note that (a) it is not clear whether the Horologium association
    really forms a tight kinematic group, because Torres et al.
    \cite{Torres00} adjusted unknown parallaxes of supposed members
    in order to minimize the velocity dispersion, and that (b) according
    to its location in the HR diagram (see Fig.~4 in Wichmann et al.~2003),
    the age of HD\,35850 is in between that of the $\beta$\,Pic 
    association ($20 \pm 10$\,Myr
    according to Barrado y Navascu\'es et al.~1999) and Per OB3, and
    within errors consistent with UCL/LCC,
    Per OB3, or $\beta$\,Pic, so its association with $\beta$ Pic is no
    more likely than our proposed 'common origin' scenario may be.

  While we clearly cannot rule out that some of these stars have
  formed in unrelated SFRs, we wish to explore here the consequences
  of the assumption of a common origin. 
  From Fig.~\ref{FigOne} we can see that at $t = -15$\,Myr, 
  the error ellipses of all of our ZAMS stars except
  HD\,77407 and HD\,129333 are located well within the 2$\sigma$ radius
  of UCL in the ($\xi$, $\eta$) plane.
  We have checked the consistency of our results by backtracking
  the motion in the $\zeta$-direction. 
  We find that at
  $t = -15$\,Myr
  the error ellipses of all our ZAMS stars overlap with both the
  LCC and UCL associations.

  de Zeeuw et al. \cite{deZeeuw99} have found 66 B-type stars in UCL,
  and 42 in LCC. Their estimated field star contamination  is 5 -- 12 (UCL) 
  and 4 -- 7 (LCC), respectively. With a normal IMF 
  (e.g. Kroupa et al.~1993), these numbers translate into
  some 200 (UCL) and 141 (LCC) G-type stars in the mass range of
  the field ZAMS stars found by us. Clearly, only a tiny fraction of
  these stars need to escape into the field to explain our results. 

  For the Per OB3 association, we find that its trajectory crosses those of 
  our 10 ZAMS stars at about $t = -50$\,Myr, i.e. the age of Per OB3. Again,
  we have checked the $\zeta$-motion for consistency.
  With $\approx 30$ B-type stars, Per OB3 is also large enough to explain
  our findings. However, Per OB3 is spatially rather compact,
  and only 3 -- 4 of our stars' error ellipses overlap with it.
  Thus, from the point of view of kinematics, we clearly prefer an LCC
  origin for our lithium excess stars.
 
  The age of the ten field ZAMS stars from our survey can only approximately
  be determined. From their lithium equivalent widths and X-ray 
  activity, we
  conclude that they are younger than the Pleiades, i.e, $<$  70 Myrs.
  In addition,
  it has been shown by Wielen \cite{Wi77} that random fluctuations
  in the local density will increase the velocity distribution with
  time. From Table 2 in his paper, we can infer an upper limit of
  of 50 -- 100\,Myr (Wichmann et al.~2003).  From their positions in
  the HR diagram (see Fig.~4 in Wichmann et al.~2003), we infer
  a lower limit of about 20 -- 30\,Myr, while ages of 50 Myrs and somewhat
  older are preferred.  We note that the absolute calibration of the
  evolutionary tracks may not be without any problems; also, the evolutionary
  tracks are computed from models without rotation, while our stars are fast 
  rotators (Wichmann et al.~2003).  At any rate, from the HR diagram's
  point of view Per OB3 would be the preferred parent association.

  We thus conclude that we cannot uniquely identify a parent association
  for our ZAMS stars.  From the point of view of
  the kinematics, the LCC/UCL associations are ideal, but with respect to
  age they may be too young. On the other hand,
  the trajectories of Per OB3 and our stars intersect at a more convenient
  age, but our backtracking shows that our stars are dispersed over a much
  large area, and only 3 -- 4 of them can be reconciled with the location
  and extent of Per OB3 at $t = -50$\,Myr.  

  In Fig. 1 -- 2 the current extents of the LCC, UCL, and Per OB3 associations
  have been
  plotted. Naturally, these need not coincide with the past extents.
  This problem has been investigated by Asiain et al. \cite{As99b}, who
  denote the Sco-Cen association as a moving group B1.
  Studying a spatially concentrated sub-group of the Sco-Cen association,
  they find that in $\xi$ and
  $\eta$, the minimum radius in the past has been about two thirds of the 
  current radius, while in $\zeta$ it has been equal to the current
  radius. If these results are applicable to the individual associations
  discussed here, our conclusions are not affected.

  If one assumes that these associations are surrounded by an isotropic
  spherical halo of dispersed stars, then from the 10 G-type stars
  in the solar vicinity we estimate the mass of that halo to
  about half of the stellar mass of the associations themselves (as derived
  from the number of B-type stars). However, both from observations
  and from theory (e.g. Wielen 1977) it is known that isotropic dispersion
  (in the sense of equipartition of energy) 
  will not lead to an isotropy of the velocity dispersions.
  In particular, the velocity dispersion
  in $W$ is smallest, because on average half of the energy gained by
  pertubations is in potential energy. 
  Therefore the shape of such a halo would be flattened along $\zeta$,
  and the amount of mass required in this haloe would be smaller.

  The associations considered here (Per OB3, and the LCC/UCL associations)
  are not gravitationally bound (de Zeeuw et al.~1999), and 
  therefore it is inescapable that stars from these associations will
  diffuse into the field. We argue that the group of field ZAMS stars
  we have found in our survey represents the ongoing dissolution of an
  OB association into the field.

  Given the similar trajectories, it may be that some of the 10 stars from
  our survey originate from Per OB3, and others from the UCL/LCC 
  associations. This may not represent that much of a difference, after all,
  because the trajectories of Per OB3 and the UCL/LCC complex intersect at
  $\approx -50$\,Myr -- corresponding to the age of Per OB3 --, so we
  speculate that the Per OB3 and LCC/UCL associations 
  may have formed out of the same molecular cloud.

  If the scenario sketched in this paper is correct, there must be many more
  young low-mass stars and brown dwarfs in the immediate vicinity of the Sun,
  independent of whether they originated in the Per OB3 or
  LCC/UCL associations.   It is a challenge to find them.

%

\section*{acknowledgements}
        This project has been supported by grants from the
        Deutsche Forschungsgemeinschaft (DFG Schwerpunktprogramm 
        `Physics of star formation'). 
	We thank the referee, Prof. R. Wielen, for his suggestions to
	improve the paper.
	This research has made use of the SIMBAD 
        database, operated at CDS, Strasbourg, France

%


\begin{thebibliography}{}

   \bibitem[1999a]{As99a}
	Asiain R., Figueras F., Torra J., Chen B. 1999,
	A\&A 341, 427

   \bibitem[1999b]{As99b}
	Asiain R., Figueras F., Torra J. 1999,
	A\&A 350, 434

   \bibitem[1984]{Ba84}
	Bahcall J.N. 1984, ApJ 276, 169
   
   \bibitem[1999]{Barrado99}
	Barrado y Navascu\'es D., Stauffer J.R., Song I., Caillault J.-P. 1999,
	ApJ 520, L123
   
   \bibitem[2002]{BMC02}
	Ben\'{\i}tez N., Ma\'iz-Apell\'aniz J., Canelles M. 2002, 
	Physical Review Letters 88, 81101 
   
   \bibitem[1964]{Bl64b}
	Blaauw A. 1964, ARA\&A 2, 213
   
   \bibitem[1942]{Ch42}
	Chandrasekhar S. 1942, Principles of Stellar Dynamics, Chicago,
	University of Chicago Press
   
   \bibitem[2001]{Chen01}
        Chen B., Stoughton C., Smith J.A. et al. 2001, ApJ 553, 184

   \bibitem[1999]{dB99}
	de Bruijne J.H.J. 1999, MNRAS 306, 181
   
   \bibitem[1999]{deZeeuw99} 
	de Zeeuw P. T., Hoogerwerf R., de Bruijne J. H. J., Brown A. G. A., 
	Blaauw A. 1999, AJ 117, 354

   \bibitem[1997]{FW97} 
	Feast M., Whitelock P. 1997, MNRAS 291, 683

   \bibitem[1999]{HA99} 
	Hoogerwerf R., Aguilar L.A. 1999, MNRAS 306, 394

   \bibitem[2001]{Ho01} 
	Hoogerwerf R., de Bruijne J. H. J., de Zeeuw P. T. 2001, A\&A, 365, 49

   \bibitem[1993]{Kroupa93} 
	Kroupa, P., Tout C.~A., Gilmore G.\ 1993, 
	MNRAS 262, 545

   \bibitem[1936]{Lb36}
	Lindblad B. 1936, Stockholm Ann. 12, No. 4, 3

   \bibitem[1942]{Lb37}
	Lindblad B. 1942, Stockholm Ann. 14, No. 1, 3

   \bibitem[2001]{MA01}
	Ma\'{\i}z-Apell\'aniz J. 2001, ApJL 560, 83

   \bibitem[2002]{Ma02}
	Mamajek E.E., Meyer M.R., Liebert J. 2002, AJ 124, 1617

   \bibitem[1968]{Mi68}
	Mihalas D., Routy P.M. 1968, Galactic Dynamics,
	Freeman, San Francisco

   \bibitem[2000]{Torres00}
	Torres C.A.O., da Silva L., Quast G.R., de la Reza R., 
	Jilinski E. 2000, AJ 120, 1410

   \bibitem[2003]{Wi02} Wichmann R., Schmitt J.H.M.M., Hubrig S. 2003,
        A\&A in press
 
   \bibitem[1977]{Wi77} Wielen R. 1977, A\&A 60, 263

   \bibitem[2001]{Zuckerman01} Zuckerman B., Inseok S., Bessel M.S.,
	Webb R.A. 2001, ApJ 562, L87


\end{thebibliography}
\end{document}